\documentclass{iaus}
\usepackage{graphics}

  \checkfont{eurm10}
  \iffontfound
    \IfFileExists{upmath.sty}
      {\typeout{^^JFound AMS Euler Roman fonts on the system,
                   using the 'upmath' package.^^J}%
       \usepackage{upmath}}
      {\typeout{^^JFound AMS Euler Roman fonts on the system, but you
                   dont seem to have the}%
       \typeout{'upmath' package installed. iaus.cls can take advantage
                 of these fonts,^^Jif you use 'upmath' package.^^J}%
      }
  \else
  \fi

  \checkfont{msam10}
  \iffontfound
    \IfFileExists{amssymb.sty}
      {\typeout{^^JFound AMS Symbol fonts on the system, using the
                'amssymb' package.^^J}%
       \usepackage{amssymb}%

      }{}
  \fi

  \IfFileExists{amsbsy.sty}
    {\typeout{^^JFound the 'amsbsy' package on the system, using it.^^J}%
     \usepackage{amsbsy}}
    {}

\newsavebox{\astrutbox}
\sbox{\astrutbox}{\rule[-5pt]{0pt}{20pt}}

\title[The Interplay among Black Holes, Stars and ISM in Galactic 
       Nuclei]{Iron Abundance Diagnostics in High-Redshift QSOs}

\author[M.R. Corbin {\it et al.\/}]%
{Michael R. Corbin,$^1$
Kirk T. Korista,$^2$ Nalaka Kodituwakku$^2$\break
\and Wolfram Freudling$^3$}

\affiliation{$^1$Dept. of Physics $\&$ Astronomy, Arizona State University,
P.O. Box 871504, Tempe, AZ 85287-1504, U.S.A, email: Michael.Corbin@asu.edu 
\\[\affilskip]
$^2$Dept. of Physics, Western Michigan University, 1903 W. Michigan Ave.
Kalamazoo, MI 49008-5252,U.S.A., email: kirk.korista@wmich.edu
\\[\affilskip]
$^3$Space Telescope - European Coordinating Facility, ESO, Karl-Schwarzchild Strasse 
2, D-85748 Garching bei M$\ddot u$nchen,  Germany: email: wfreudli@eso.org} 
\pubyear{2004}
\volume{222}
\pagerange{1--8}
\date{?? and in revised form ??}
\jname{The Interplay among Black Holes, Stars and ISM \\in Galactic Nuclei}
\editors{Th. Storchi Bergmann, L.C. Ho \& H.R. Schmitt, eds.}
\begin{document}

\maketitle

\begin{abstract}
The abundance of $\alpha$-process elements such as magnesium and carbon relative to iron measured from the broad emission lines of QSOs can serve as a diagnostic of the star formation and chemical enrichment histories of their host galaxies.  We investigate the relationship between Fe/Mg and Fe/C abundance ratios and the resulting Fe II / Mg II $\lambda2800$ and Fe II / 1900$\AA$-blend flux ratios, both of which have been measured in QSOs out to {\it z} $\approx$ 6.  Using a galactic chemical evolution model based on a starburst in a giant elliptical galaxy, we find that these flux ratios are good tracers of the chemical enrichment of the nuclei. However, the values of these ratios measured in objects at {\it z} $\approx$ 6 suggest that iron enrichment has occurred more rapidly in these objects than predicted by the assumed elliptical starburst model, assuming currently favored cosmologies.  This in turn suggests that refinements to the galactic chemical evolution models are needed.
\end{abstract}

\firstsection 
\section{Introduction}
Strong Fe II emission in the UV bump at 2500$\AA$ has now been detected in QSOs out to $\it z$ $\approx$ 6 (Freudling, Corbin \& Korista 2003; Barth et al. 2004). The strengths of the Fe II 2500$\AA$ feature, Mg II $\lambda$2800, and the blend of C III] $\lambda$1909, Al III $\lambda$1860 and Si III] $\lambda$1892 lines at approximately 1900$\AA$ in these objects and the associated line ratios are quite similar to those measured in QSOs at lower redshifts.  While the iron emission processes in the broad line region (BLR) remain incompletely understood, the measured line ratios suggest that the BLRs in these objects have metallicities of several times the solar value (see, e.g., Hamann et al. 2002; Dietrich et al. 2003).  Motivated in particular by the results for objects at $\it z$ $\approx$ 6, which under the currently favored cosmology of $\it h$ = 0.7, $\Omega_M$ = 0.3, $\Omega_\Lambda$ = 0.7 corresponds to an age of the universe of approximately 0.9 Gyr, we seek to refine these metallicity estimates using recent Fe II atom and galactic chemical enrichment (GCE) models.

\section{Line Flux vs. Elemental Abundance}
Using Ferland's code CLOUDY, we compute several grids of photoionization models, each spanning several orders of magnitude in gas density and ionizing flux.  Each grid assumes a set of metal abundances corresponding to the giant elliptical starburst model used by Hamann \& Ferland (1999).  The iron enrichment in this GCE model arises primarily from Type Ia supernovae. We find the relations between the measured Fe II 2500$\AA$ bump / Mg II $\lambda$2800 flux ratio and the associated Fe/Mg abundance ratio using both a simplified Fe II atom model and a more recent 371-level model to be log(FeII/MgII) $\approx$ -0.13 + 0.4log(Fe/Mg) andlog(FeII/MgII) $\approx$ -0.13 + 0.6log(Fe/Mg), respectively.  Importantly, in both cases the relation is not linear, as has been assumed in some studies (e.g. Yoshii, Tsujimoto \& Kawara 1998), as the Fe II / Mg II flux ratio scales only moderately with the Fe/Mg abundance ratio.   

We additionally investigate the scaling of the Fe/C abundance with the Fe II / 1900$\AA$-blend flux ratio, and find similar relations. This supports the use of this ratio as a metallicity diagnostic in objects in which Mg II $\lambda$2800 cannot be measured (c.f. Freudling et al. 2003). 
\section{Time Evolution}
Using the relations derived between the Fe II 2500$\AA$ bump / Mg II $\lambda$2800 and Fe II 2500$\AA$ bump / 1900$\AA$-blend flux ratios and the respective cabundance ratios for the case of the 371-leve Fe II atom, we use the Hamann \& Ferland (1999) giant elliptical GCE model to predict the values of these flux ratios as a function of time, over a time scale of approximately 6 Gyr from the initial starburst.  Interestingly, the values of these ratios do not reach those measured for $\it z$ $\approx$ 6 QSO (see the references in $\S$ 1) until more than 1 Gyr after the onset of star formation, which is inconsitent with the age of the universe at such redshifts under a $\it h$ = 0.7, $\Omega_M$ = 0.3, $\Omega_\Lambda$ = 0.7 cosmology.  Assuming this cosmology to be correct, this indicates that the GCE models need to be refined, allowing for more rapid iron enrichment. One such refinement may be in the contribution of SNe Ia to the overall iron yield; Venkatesan, Schneider \& Ferrara (2004) find that SNe Ia are not strongly required in order to produce the inferred iron abundances of high-redshift QSOs.   

\section{Conclusions}
The detection of strong iron emission and apparent super-solar metallicities in QSOs at $\it z$ $\approx$ 6 places important constraints on models of galactic chemical enrichment and formation, as well as the redshifts at which the first stars in the universe formed.  Using the most recent Fe II atom models, we find that important diagnostic flux ratios including Fe II 2500$\AA$ / Mg II $\lambda$2800 scale more gradually with the abundance ratios of these elements than some studies have assumed.  Under currently favored cosmologies, we additionally find an inconsistency between the values of these ratios and those predicted by recent galactic chemical evolution models, in that the values measured in objects at $\it z$ $\approx$ 6 are higher than the model predictions.  This indicates that refinements to these models are needed, perhaps in the amount of contribution by SNe Ia to the iron yield.    

\end{document}